\documentstyle[11pt,paspconf,epsf]{article}

\def \farcs{\hbox{$.\!\!^{\prime\prime}$}}

\begin{document}

\title{Microlensing \& Scintillation of Gravitationally Lensed
       Compact Radio Sources: Evidence for MACHOs?}

\author{L.V.E. Koopmans$^1$ \& A.G. de Bruyn$^{2,1}$ }

\affil{$^1$Kapteyn Astronomical Institute, P.O.Box 800, NL-9700 AV 
       $^{~}$Groningen, the Netherlands\\
       $^2$NFRA, P.O.Box.2, NL-7990 AA, Dwingeloo, The Netherlands}

\begin{abstract}
We present the first unambiguous case of external variability of an
extra--galactic radio source, the CLASS gravitational lens B1600+434,
consisting of two images lensed by an edge--on disk galaxy.  The VLA
8.5--GHz difference light curve of the lens images shows external
variability at the 14.6--$\sigma$ confidence level. Although the
current single--frequency VLA observations cannot conclusively
exclude scintillation, several lines of evidence show that it does not
dominate the short--term variability. This is supported by an ongoing
three--frequency WSRT campaign. If scintillation is the dominant
source of this external variability, it would require very different
properties of the Galactic ionized ISM towards the two lens images.
Microlensing of a superluminal jet component can explain both
the rms and time--scale of variability, but requires the halo of the
lens galaxy to be filled with $\ga$0.5--M$_\odot$ MACHOs. The current
data appears to support microlensing as the dominant source of the
observed external variability.
\end{abstract}


\section{Introduction}

Gravitationally lensed compact radio sources have many astrophysical
and cosmological applications, amongst which the determination of a
time--delay to constrain the Hubble constant is best known (Refsdal
1964). Once a time--delay has been determined from the image light
curves, they can be subtracted to see if any externally induced
variability is present. External variability in compact radio sources
can be the result of either scintillation by the Galactic ionized ISM
(Rickett 1990) or microlensing by compact objects in the lens galaxy
(Chang \& Refsdal 1979).  Unambiguous external variability thus offers
the opportunity to study either the ionized medium towards the lens
images or the mass function of compact objects in high--z
galaxies. Here, we discuss both possibilities as the source of the
external variability observed in the CLASS gravitational lens
B1600+434 (Koopmans \& de Bruyn 1999 [KB99]).

\section{Observations}

\begin{figure}
\plotone{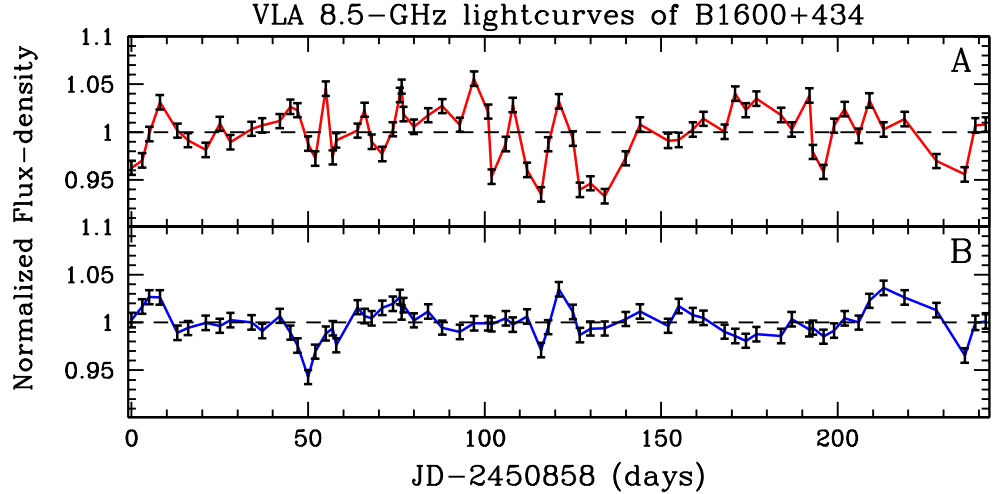}
\caption{The normalized light curves of B1600+434 A (upper) and B
(lower), corrected for a long--term gradient. The error on each
light--curve epoch is 0.7 to 0.8\% (KB99).}
\label{fig-1}
\end{figure}

\begin{figure}
\plotone{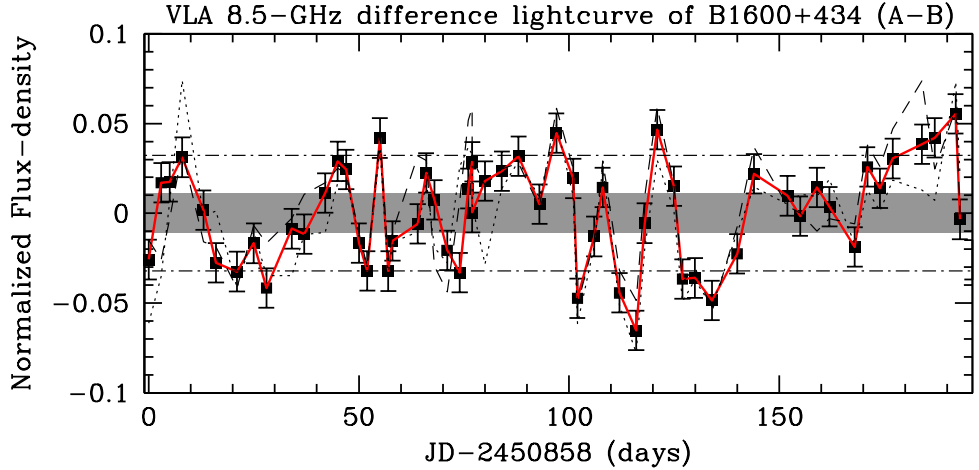}
\caption{The normalized difference light curve between the two
lens images. The shaded region indicates the expected 1--$\sigma$
(1.1\%) region if all variability were due to measurement errors. The
dash--dotted lines indicate the observed rms variability of 3.2\%. The
dotted and dashed curves indicate the normalized difference curves for
a time delay of 41 and 52 days, respectively (KB99).}
\label{fig-2}
\end{figure}

B1600+434 was observed from February to October 1998 with the VLA at
8.5--GHz, in order to determine a time delay between the two lens
images (Koopmans et al. 1999 [KBXF99]).  In Fig.\ref{fig-1}, the
normalized VLA 8.5--GHz light curves of both lens images are shown,
created by subtracting a linear fit and subsequently normalizing them
to unity (KBXF99). The resulting normalized light curves have rms
variabilities of 2.8\% and 1.6\% for images A and B, respectively.
Subsequently, the observed linearly--interpolated light curve of
image~B was subtracted from the light curve of image~A, after having
corrected it for a flux density ratio of 1.212 and a time delay of 47
days (KBXF99).  The normalized difference light curve
(Fig.\ref{fig-2}) has an rms variability of 3.2\%, which is
inconsistent with a flat difference light curve at the 14.6--$\sigma$
confidence level. A time delay different from 47 days only increases
the confidence level, as a result of using a minimum--dispersion
method to determine the time delay (KBXF99).

\section{Analysis}

Below we investigate both scintillation and microlensing, as possible
causes of the external variability in the lens images of B1600+434.
We regard extreme scattering (Fiedler 1987) as unlikely and will
not further discuss it here. The reader is referred to KB99 for a full
discussion of all the details.

\subsection{Scintillation}

Scintillation, caused by the Galactic ionized ISM, can explain both
large amplitude variability in very compact extra--galactic radio
sources and lower amplitude `flicker' of more extended sources
(Rickett 1990).  In B1600+434--A and B, we observe several percent rms
variability, possibly consistent with `flickering'. The time scale is
harder to quantify, but it appears that we see variability time scales
of several days up to several weeks (Figs \ref{fig-1}--\ref{fig-2}).

Let us summarize the arguments against scintillation: (i)
The modulation index between the lens images differs, even though one
looks at the same background source. This requires either
significantly different scattering measures (SM$_{\rm A}$/SM$_{\rm
B}$=3.1) over a scale comparable to the image separation (1\farcs4) or
an image size ratio $\Delta\theta_{\rm B}$/$\Delta\theta_{\rm
A}$=1.75, with $\Delta\theta_{\rm A}$$\approx$60 $\mu$as.  (ii)
Ongoing multi--frequency WSRT total--flux observations ($\ga$6 months)
shows a decrease in the modulation index with wavelength
($m_{6}/m_{21}$$\sim$2), whereas the opposite trend is expected for
refractive scintillation. If one would attribute all short--term
variability to intrinsic source variations, it becomes very difficult
to reconcile this with the fact that most short--term variability at
3.6 cm (8.5--GHz) is external and not intrinsic. The logical
conclusion would be that most short--term variability seen in the WSRT
6--cm light curves is also of external origin with a modulation index
smaller than that at 21 cm. (iii) A modulation index of a few percent
corresponds to a variability time scale of less than 1--2 days for
weak and strong refractive scattering, whereas variability over much
longer scales is present.

\subsection{Microlensing}

For several reasons, microlensing is a plausible source of the
external variability of B1600+434: (i) Because B1600+434 is multiply
imaged, the lens images pass through a fore--ground lens galaxy with
high microlensing optical depths, i.e.~$\tau_{\rm A}$$\approx$0.2 and
$\tau_{\rm B}$$\approx$0.9 (KB99). Combined with the fact that many
core--dominated flat--spectrum radio source have superluminal jet
components (Vermeulen \& Cohen 1994), makes it probable that
microlensing variability is occurring at some level on time scales of
weeks to months. (ii) From microlensing simulations (Wambsganss 1990),
we find that a static core, which does not vary on short time scales
due to microlensing (i.e. $v_{\rm core}$$\ll$$v_{\rm knot}$), plus a
single compact ($\Delta\theta_{\rm knot}$=2--5 $\mu$as) superluminal
($\beta_{\rm app}$$\ga$9) jet component that contains 5--11\% of the
source flux, can explain the observed rms and time scale of
variability in the lens images.  For image~A, however, we require an
average mass of compact objects $\ga$0.5--M$_\odot$ to explain its
higher rms variability (even though $\tau_{\rm A}$$<$$\tau_{\rm B}$).
A smaller MACHO mass would result in less variability, due to the
convolution of a finer magnification pattern with the source
structure. (iii) The decrease in rms variability ($m_{\lambda}$) with
wavelength (i.e. $m_{\rm 21}$$<$$m_{\rm 6}$), seen in the WSRT
observations (Sect.3.1), is in agreement with a microlensed synchrotron
self--absorbed superluminal jet component that grows proportional with
wavelength, in which case one expects $m_{6}/m_{21}$$\approx$2--3,
comparable to the observed ratio (Sect.3.1).

\section{Conclusions}

We have presented the first unambiguous case of external variability
of an extra--galactic radio source, the CLASS gravitational lens
B1600+434.  Several lines of evidence point to microlensing of a
superluminal jet component and not scintillation as the dominant
source of this external variability. This implies the presence of a
population of MACHOs in the halo around the edge-on disk galaxy with
masses $\ga$0.5--M$_\odot$. The fact that these compact ($\la$5
$\mu$as) structures do not seem the scintillate remains a problem to
be resolved, but might be the result of the assumed simple source
structure and/or the idealized mass functions of compact objects.
Ongoing multi--frequency WSRT and VLA observing campaigns should
further tighten the constraints on this exciting possibility to 
learn something about the dark matter properties at high redshifts.

\acknowledgments
The authors thank the Cosmic Lens All--Sky Survey (CLASS)
collaboration, without which this work would not have been possible.
LVEK and AGdeB acknowledge the support from an NWO program subsidy
(grant number 781-76-101). This research was supported in part by the
European Commission, TMR Programme, Research Network Contract
ERBFMRXCT96-0034 `CERES'.

\end{document}